\documentstyle[aps,multicol,epsfig]{revtex}

\def\e{{\epsilon}}
\def\k{{ {\bf k} }}
\def\p{{ {\bf p} }}
\def\q{{ {\bf q} }}

\def\w{{\omega}}

\begin{document}
\draft

\title{
Reply to Comment on "Theory of Hall Effect and Electrical Transport in 
High-Tc Cuprates: Effects of Antiferromagnetic Spin Fluctuations"
}

\author{
H. Kontani$^{1}$ and 
K. Kanki$^{2}$ 
}

\address{
$^1$Department of Physics, Saitama University,
255 Shimo-Okubo, Urawa-city, 338-8570, Japan.
\\
$^2$College of Integrated Arts and Sciences, Osaka 
Prefecture University, Sakai 599-8531, Japan.
}

\date{\today}

\maketitle


\begin{multicols}{2}
\narrowtext

Recently, we have studied the magnetotransport phenomena
in high-Tc cuprates and in $\kappa$-BEDT-TTF organic superconductors
on the basis of the Fermi liquid theory
 \cite{Hall,MR-HT}.
Based on the 'exact' expressions 
for the Hall coefficient ($R_{\rm H}$)
 \cite{Kohno}
and for magnetoresistance (MR)
 \cite{MR},
we studied their behavior in nearly antiferromagnetic (AF)
Fermi liquid state.
We take care to satisfy the Ward identity
 \cite{Baym}
in order not to violate the conserving laws
and several rigorous relations in the Fermi liquid.
In this article, we explain the essence of our works
and point out the mistakes in the commet by O. Narikiyo
\cite{Nari}.

(i) {\it Yamada-Yosida theory} : \ 
Based on the Fermi liquid theory,
Yamada and Yosida proved exactly that 
the resistance becomes 'zero' even at finite temperatures,
if there are no Umklapp processes
 \cite{Yamada}.
This rigorous result is reproduced within the fluctuation-exchange
(FLEX) approximation with all the vertex corrections required by
the Ward identity;
one Maki-Thompson (MT) term and two Aslamazov-Larkin (AL) terms
as shown in Fig.\ref{fig:T22}.
It is proved in Appendix D of Ref.\cite{MR}, 
by using the fact that the Bethe-Sapleter equation
in the FLEX approximation is equial to the exact one,
eq.~(6$\cdot$13) or (6$\cdot$17) of Ref.\cite{Yamada},
except for the kernel function; $\Delta_0(k,k';k'+q,k-q)$.
This fact means that 
the analysis of the conductivity with Umklapp processes in 
Ref.\cite{Yamada} is also reproduced by the FLEX.
In conclusion, the related criticism in Ref. \cite{Nari} is false
as stated above.
\begin{figure}
\begin{center}
\epsfig{file=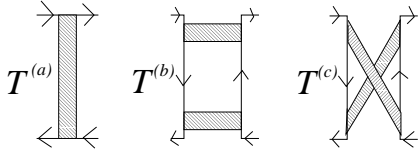,width=5.5cm}
\end{center}
\caption{
One MT-term (${\cal T}^{(a)}$) and two AL-terms (${\cal T}^{(b,c)}$).
The box represents the spin fluctuation
in the FLEX.
}
  \label{fig:T22}
\end{figure}

\begin{figure}
\begin{center}
\epsfig{file=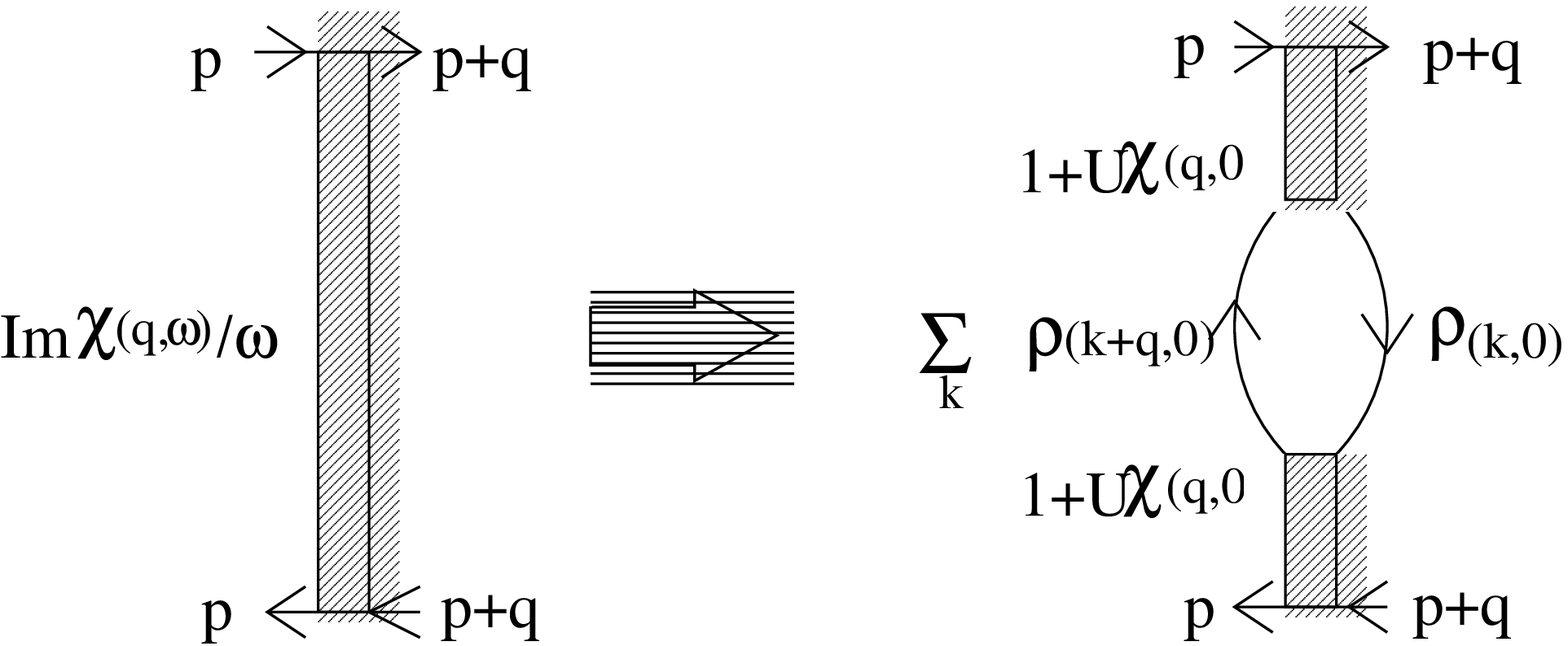,width=6.5cm}
\end{center}
\caption{
${\cal T}_{22}^I(\p,\p+\q)$  
coming from the MT-term.
}
  \label{fig:MT}
\end{figure}

(ii) {\it ${\cal T}_{22}^I$ in the FLEX} : \ 
According to Eliashberg
 \cite{Eliashberg},
${\cal T}_{22}^I(p,p')$ 
is given by the difference 
across the branch cuts, so it is 'pure imaginary'.
As is known,
the imaginary part of a four-point vertex at finite energies
should contain (at least) two $\rho_\k(0)$'s in it
 \cite{AGD}.
This fact is overlooked in Ref. \cite{Nari}.
As for the MT-term,
${\cal T}_{22}^I(p,p+q)$
in the FLEX approximation is
proportional to $(U^2T^2){\rm Im}\chi_\q(\w)/\w|_{\w=0}$
if charge fluctuations are negligible
 \cite{Hall}.
$\chi_\q(\w)= \chi_\q^0(\w)/(1-U\chi_\q^0(\w))$ 
is the spin susceptibility,
and 
$\chi_\q^0(\w) = -T\sum_{\e,\k} G_\k(\e)G_{\k+\q}(\e+\w)$.
We can show that 
\begin{eqnarray}
& &{\rm Im}\chi_\q(\w)/\w|_{\w=0}
 = (1+U\chi_\q(0))^2 \cdot {\rm Im}\chi_\q^0(\w)/\w|_{\w=0}
 \nonumber \\
& &  \ \ \ \ \ \ \ \ \ 
 = (1+U\chi_\q(0))^2 \cdot \pi\sum_\k 
     \rho_\k(0) \rho_{\k+\q}(0).
\end{eqnarray}
Thus, ${\cal T}_{22}$ coming from the MT-term 
(${\cal T}^{(a)}$ in Fig.\ref{fig:T22})
is shown in the r.h.s of Fig.\ref{fig:MT},
which is compatible with the Fermi liquid theory
because $\rho_\k(\w\!=\!0)$ represents the 'coherent part'.
Thus, the criticism against the MT-term of the FLEX approximation
in Ref.\cite{Nari} is incorrect.
This term together with ${\cal T}^{(b,c)}$ in Fig.\ref{fig:T22}
form the three diagrams in Fig.5 of Ref.\cite{Yamada},
so Yamada-Yosida theory is satisfied exactly in the FLEX
as explained in (i).

When the AF fluctuations are strong due to nesting
like in high-$T_{\rm c}$ cuprates,
which is beyond the scope of the analysis by Yamada-Yosida,
${\cal T}_{22}$ gives the additional temperature dependence
on both $R_{\rm H}$ and MR.
The mechanism, which is the main conclusion of Ref.\cite{Hall,MR-HT},
are based on the established Fermi liquid theory.

In summary,
the criticisms by Ref.\cite{Nari} 
are misunderstandings.
The famous 'seemingly' non-Fermi liquid behavior
of the magnetotransport phenomena 
in high-Tc and in $\kappa$-BEDT-TTF 
is naturally understood from the standpoint of 
the nearly AF Fermi liquid.

\vspace{-5mm}


\end{multicols}

\end{document}